\documentstyle[12pt,epsfig]{article}
\voffset     = - 0.10 in
\begin{document}
\thispagestyle{empty}


\hfill {\Large\bf \today }

\vspace*{1.0 in}

{\Large\bf
\begin{center}
On the e/h Ratio of the Electromagnetic Calorimeter
\end{center}
}

\bigskip
\bigskip

\begin{center}
{\large\bf Y.A.\ Kulchitsky}
\footnote{
E-mail: Iouri.Koultchitski@cern.ch}

\bigskip

{\sl 
Joint Institute for Nuclear Research, Dubna, Russia}
\smallskip

{\sl 
Institute of Physics, National Academy of Sciences, Minsk, Belarus}\\
\end{center}

\vspace*{\fill}

\begin{abstract}
The method of extraction of the $e/h$ ratio for electromagnetic compartment 
of combined calorimeter is suggested and the non-compensation was determined.
The results agree with the Monte Carlo prediction and results of the
weighting method for electromagnetic compartment of combined calorimeter.
The new easy method of a hadronic energy reconstruction for a combined 
calori\-meter is also suggested.
The proposed methods can be used for combined calorimeter, which is being
designed to perform energy measurement in a next-generation high energy 
collider experiment like ATLAS at LHC. 
\vskip 5mm 
\noindent
{\bf Codes PACS:}
29.40.Vj, 
29.40.Mc, 
29.85.+c., 
\\
{\bf Keywords:} 
Calorimetry, 
Shower Counters, 
Combined Calorimeter,
Compensation,
Energy Measurement, 
Computer Data Analysis.
\end{abstract}
\newpage

\section{Introduction}

The future experiment ATLAS \cite{ATL94,ATL99V1} 
at the Large Hadron Collider
(CERN) will include a combined calorimeter \cite{CAL96}
with in the central region the two separate units: 
the liquid-argon electromagnetic ca\-lorimeter \cite{LARG96}
and the iron-scintillating hadronic calorimeter 
\cite{TILECAL96,HS98,CALOR99,nima443}. 

For many tasks of calorimetry it is necessary to know a non-compensa\-tion of 
combined calorimeter compartments. 
As to the hadronic calorimeter there is the detailed information about the 
$e/h$ ratio presented in \cite{TILECAL96,budagov96-72,kulchitsky99}.
But as to the electromagnetic calorimeter \cite{ccARGON} reliable information 
practically absent.

The aim of the present work is to develop the method for the determination 
of the electromagnetic compartment non-compensation and compares 
results of this method with results of weighting method 
\cite{stipcevic93,gingrich95} and Monte Carlo prediction \cite{wigmans91} 
for the same calorimeter.
The new method of an energy reconstruction for combined calorimeter is 
also presented.
For detailed understanding of performance of the future calorimetry 
the combined calorimeter setup has been made consisting of the liquid-argon
electromagnetic calorimeter inside the cryostat and downstream the 
iron-scintillating hadronic calorimeter \cite{comb94,comb96,cobal98}.

\section{Method}

The response, $R$, of a calorimeter to a hadronic shower is the sum of 
the contributions from the electromagnetic, $E_e$, and hadronic, $E_h$,  
parts of the hadronic shower energy, $E = E_e + E_h$, 
\cite{groom89}
\begin{equation}
  R = e\cdot E_e + h\cdot E_h \ , 
\label{ev9}
\end{equation}
where $e$ ($h$) is the energy independent coefficient of transformation
electromagnetic (hadronic) part of a shower energy to response.
Therefor an incident energy is
\begin{equation}
  E = (1/e) \cdot (e/\pi) \cdot R \ , 
\label{ev16}
\end{equation}
where
\begin{equation}
        \frac{e}{\pi}=\frac{e/h}{1+(e/h-1)\cdot k\cdot\ln{(E)}} \ , 
\label{ev10}
\end{equation}
$f_{\pi^0} = k \cdot \ln{(E)} = E_{e}/E $ is a fraction of electromagnetic 
energy.
In the case of a combined calorimeter the incident energy is deposited into 
an electromagnetic compartment, $E_{em}$, into a hadronic compartment, 
$E_{had}$, and into a dead material between the two calorimeters, $E_{dm}$.
Using relation (\ref{ev16}) the following expression has been obtained:
\begin{equation}
  E = E_{em} + E_{dm} + E_{had} = 
      \frac{1}{e_{em}}\Biggl(\frac{e}{\pi}\Biggr)_{em} R_{em} +E_{dm}
    + \frac{1}{e_{had}}\Biggl(\frac{e}{\pi}\Biggr)_{had} R_{had}\ ,
\label{ev7}
\end{equation}
where $R_{em}$ ($R_{had}$) is response of a electromagnetic (hadronic) 
calorimeter compartment, $1/e_{em}$ \cite{comb96,kulchitsky98-336}
and $1/e_{had}$ \cite{comb96} are the energy calibration constant for 
electromagnetic and hadronic calorimeter.

The Eq.\ (\ref{ev7}) is the basic formula for the new, non-parametrical, 
method of a hadronic energy reconstruction for a combined calorimeter. 
This method does not require the determination of any parameters by a 
minimisation technique and uses known $e/h$ ratios and electron 
calibration constants.
In the right side of the Eq.\ (\ref{ev7}) an energy is under a logarithmic 
function therefore for achievement of convergence with an accuracy of 
$\approx 1\%$ is sufficiently only the first approximation.
The obtained reconstruction of the mean values of energies is within 
$\pm 1\%$ and this accuracy can be compared with results from 
Ref.\ \cite{cavalli96,comb96}. 
The fractional energy resolution is comparable with the benchmark method 
result \cite{comb96}.
The method can be used for the fast energy reconstruction in the trigger.

From expression (\ref{ev7}) the value of the $(e/\pi)_{em}$ ratio can be 
obtained
\begin{equation}
\Biggl(\frac{e}{\pi}\Biggr)_{em}=
        \frac{E_{beam}-E_{dm}-E_{had}}{R_{em}\cdot (1/e_{em})}\ . 
\label{ev1}
\end{equation}
The $(e/h)_{em}$ ratio can be inferred from (\ref{ev10}), where $E$ is
the beam energy.
For calculation of the $E_{had}$ the value $(e/h)_{had}$ 
\cite{budagov96-72} was used and $E$ in the (\ref{ev10}) is the energy 
deposited in the hadronic calorimeter, $k = 0.11$ \cite{wigmans91}.
The term $E_{dm}$ is taken similar to \cite{comb96,comb94}:
$E_{dm}= (1/e_{dm})\cdot\sqrt{E_{em, l}\cdot E_{had, f}}$, where
$E_{em, l}$ is an energy released in a last depth of an electromagnetic 
calorimeter and $E_{had, f}$ is an energy released in a first depth of a 
hadronic calorimeter.
The validity of this approximation has been tested by the experimental study 
\cite{cobal98,comb96} and by the Monte Carlo simulation 
\cite{efthym96,bosman99}.

\section{Results}

\begin{figure*}[tbph]
\begin{center}   
\mbox{\epsfig{figure=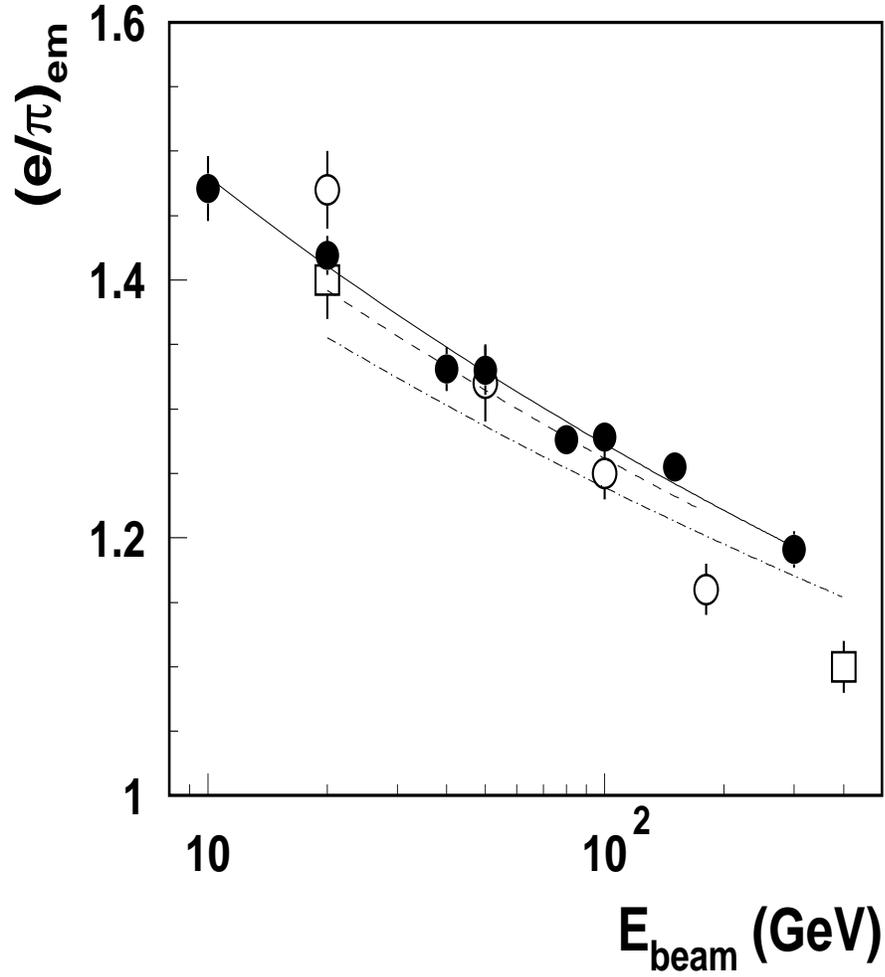,width=0.95\textwidth,height=0.8\textheight}}
\end{center}
       \caption{
        The $(e/\pi)_{em}$ ratios as a function of the beam energy for this
        method (black circles) and for weighting method (open circles for 
      Ref.\ \cite{stipcevic93} and open squares for Ref.\ \cite{gingrich95}).
        The lines are the result of a fit of Eq.\ (\ref{ev10}) with free 
        $e/h$ parameter and $k=0.11$: solid line is for the
        \cite{kulchitsky99-303} data, dashed line
        is for the \cite{stipcevic93} data and dash-doted line is 
        for the \cite{gingrich95} data.
      \label{fv3}}
\end{figure*}

The mean values of the $(e/\pi)_{em}$ distributions, derived by (\ref{ev1}) 
and extracting by fitting in the $\pm2\sigma$ \cite{kulchitsky99-303}, 
are given in Table \ref{tv1} and 
shown in Fig.\ \ref{fv3} (black circles) as a function of the energy.
The fit of $(e/\pi)_{em}$ values by the expression (\ref{ev10}), with two 
parameters, yields $(e/h)_{em} = 1.74\pm0.04$ and $k = 0.108\pm0.004$. 
The value of parameter $k$ is in the good agreement with well known 
$0.11$ \cite{wigmans91}.
For fixed parameter $k$ the value of non-compensation is 
$(e/h)_{em}=1.77\pm0.02$.
The quoted errors are the statistical ones and obtained from the fit.
The systematic error, which is a consequence of the uncertainties in the 
input constants used in the (\ref{ev1}), is estimated to be $\pm0.04$.

\begin{table}[tbph]
\begin{center}
\caption{
The $(e / \pi)_{em}$ ratios as a function of the beam energy.}
\label{tv1}
\begin{tabular}{|c|c|c|c|}
\multicolumn{4}{l}{\mbox{~~~}}          \\
\hline
$E$&
\multicolumn{3}{c|}{$(e / \pi)_{em}$}
\\ 
\cline{2-4}
(GeV)&\cite{kulchitsky99-303}&\cite{stipcevic93}&\cite{gingrich95}\\ 
\hline
10              & $1.47\pm0.03$& --          & -- \\ 
\hline 
20              & $1.42\pm0.02$&$1.47\pm0.03$&$1.40\pm0.03$ \\ 
\hline 
40              & $1.33\pm0.02$& --          & --  \\ 
\hline 
50              & $1.33\pm0.02$&$1.32\pm0.03$& -- \\ 
\hline 
80              & $1.28\pm0.01$& --          & --  \\ 
\hline 
100             & $1.28\pm0.01$&$1.25\pm0.02$& -- \\ 
\hline 
150             & $1.26\pm0.01$& --          & -- \\ 
\hline 
180             & --           &$1.16\pm0.02$& -- \\ 
\hline 
300             & $1.19\pm0.02$&$0.96\pm0.02$& -- \\ 
\hline 
400             & --           & --          &$1.10\pm0.02$\\ 
\hline 
\end{tabular}
\end{center}
\end{table}

In the Ref.\ \cite{wigmans91} showed that the $e/h$ ratio for non-uranium
calorimeters with high-$Z$ absorber material is satisfactorily described
by the formula:
\begin{equation}
        \frac{e}{h}=\frac{e/mip}{0.41 + f_n \cdot n/mip}\ ,
\label{wig}
\end{equation}
where $f_n$ is a constant determined by the $Z$ of the absorber (for lead 
$f_n=0.12$) \cite{wigmans87,wigmans98}, 
$e/mip$ and $n/mip$ represent the calorimeter 
response to electromagnetic showers and to MeV-type neutrons, respectively.
These responses are normalised to the one for minimum ionising particles.
The Monte Carlo calculated $e/mip$ and $n/mip$ values \cite{wigmans88}
for the lead-liquid-argon electromagnetic calorimeter \cite{costa91} are 
$e/mip = 0.78$ and $n/mip < 0.5$ and leading to $e/h > 1.66$.
The measured value of the $(e/h)_{em}$ ratio agrees with this prediction.  

The formula (\ref{wig}) show that $e/mip$ is very important for 
understanding compensation in lead-liquid-argon calorimeters.
The non-compensation increase when the sampling frequency is also increased 
\cite{wigmans87}.
A large fraction of the electromagnetic energy is deposited through very 
soft electrons ($E < 1$ MeV) produced by Compton scattering or the 
photoelectric effect.
The cross sections for these processes strongly depend on $Z$ and 
practically all these photon conversions occur in the absorber material.
The range of the electrons produced in these processes is very short, 
$\sim 0.7$ mm for 1 MeV electron in lead.
Such electrons only contribute to the calorimeter signal if they are produced
near the boundary between the lead and the active material.
If the absorber material is made thinner this effective boundary layer 
becomes a larger fraction of the total absorber mass and the calorimeter 
response goes up.
This effect was predicted by EGS4 simulation \cite{wigmans87}.
It explains that predictions for the GEM \cite{barish92} accordion 
electromagnetic calorimeter (1 mm lead and 2 mm liquid-argon) are 
the $e/mip = 0.86$ and the $e/h > 1.83$.
The Monte Carlo calculations also predict that the electromagnetic response
for liquid-argon calorimeters (due to the larger $Z$ value of argon) is 
consistently large than for calorimeters with plastic-scintillator readout.
The signal from neutron ($n/mip$) suppressed with factor $0.12$ and 
the $n-p$ elastic scattering products do not contribute to the signal of 
liquid-argon  calorimeters. 
These detectors only observe the $\gamma$'s produced by inelastic neutron 
scattering and from thermal neutron capture 
\cite{wigmans87}. 

In the Refs.\ \cite{stipcevic93,gingrich95} the following definition of 
an $e/\pi$ ratio for first compartment of the combined calorimeter is 
adopted.
The estimators for pion and electron energies, respectively, are
$E = c_{em}^{\pi}\cdot R_{em} + c_{had}^{\pi}\cdot R_{had}$ and 
$E = c_{em}^e \cdot R_{em}$,            
where $R_{em}$ and $R_{had}$ are responses of elecromagnetic and hadronic 
compartments of a combined calorimeter, $c_{em}^e$ (energy 
independent within $1\%$) is the energy calibration constant for the 
electromagnetic calorimeter, $c_{em}^{\pi}$ and $c_{had}^{\pi}$ are 
weighting parameters for pions.
These parameters was find using a minimisation procedure for a energy 
resolution ($\sigma/E$) at every beam energies.
In the Ref.\ \cite{stipcevic93,gingrich95} an electron/pion ration defined as
$(e/\pi)_{em} = c_{em}^\pi/c_{em}^e$.
This definition one can find from (\ref{ev7}) for an electromagnetic 
compartment, where $c_{em}^{\pi} = 1/e_{em} \cdot (e/\pi )_{em}$ and
$1/e_{em} = c_{em}^e$.

The results of this weighting method for $(e/\pi)_{em}$ rations are 
given in Table \ref{tv1} and 
shown in Fig.\ \ref{fv3} (open circles are for \cite{stipcevic93} and open 
squares are for \cite{gingrich95}).
In the energy region $\leq 100$ GeV the \cite{kulchitsky99-303} data are in 
a good agreement with \cite{stipcevic93,gingrich95} data and in disagreement 
for energies $> 100$ GeV.
Fit of the $(e/\pi)_{em}$ values by the expression (\ref{ev10}), with two 
parameters, yields $(e/h)_{em} = 2.28\pm0.19$ and $k = 0.143\pm0.006$ for 
\cite{stipcevic93} data and $(e/h)_{em} = 1.93\pm0.13$ and 
$k = 0.135\pm0.007$ for \cite{gingrich95} data.
Note, that problematical value of $(e/\pi)_{em}=0.96\pm0.02$ at 300 GeV 
\cite{stipcevic93} is excluded from the fit. 
One can see that parameters $k$ are more bigger that its well known value
and the $(e/h)_{em}$ are bigger than our result. 
For fixed parameter $k=0.11$ the result of the fit are 
$(e/h)_{em} = 1.73\pm0.10$ for \cite{stipcevic93} data and 
$(e/h)_{em} = 1.64\pm0.18$ for \cite{gingrich95} data.
In the both cases we calculated errors of the $e/h$ taken into account 
the values of $<\chi^2>$. 
The finding $e/h$ rations are in agreement with our result within error bars.
Therefore, one can see that the weighting method leads to distortion of the 
$(e/\pi)_{em}$ ratios.

\section{Conclusions}

The method of extraction of the $e/h$ ratio for electromagnetic compartment 
of combined calorimeter is suggested and the non-compensation was determined.
The results agree with the Monte Carlo prediction and results of the
weighting method for electromagnetic compartment of combined calorimeter.
The new easy method of a hadronic energy reconstruction for a combined 
calori\-meter is also suggested.
The proposed methods can be used for combined calorimeter, which is being
designed to perform energy measurement in a next-generation high energy 
collider experiment like ATLAS at LHC. 
 
\section{Acknowledgement}
 
Author would like to thank P.\ Jenni, J.\ Budagov and M.\ Nessi for 
fruitful discussions and attention for this work.
I am grateful M.\ Kuzmin, V.\ Vinogradov, F.\ Gianotti and M.\ Cobal 
for constructive advices and fruitful discussions.



\end{document}